\documentclass[pdflatex,sn-mathphys]{sn-jnl}

\jyear{2021}%

\theoremstyle{thmstyleone}%
\theoremstyle{thmstyletwo}%

\theoremstyle{thmstylethree}%

\begin{document}

\title[Isik et al.]{Design optimization for high-performance computing using FPGA}


\author*[1]{\fnm{Murat} \sur{Isik}}\email{mci38@drexel.edu}

\author[2]{\fnm{Kayode} \sur{Inadagbo}}\email{kayodeinadagbo@gmail.com}

\author[3]{\fnm{Hakan} \sur{Aktas}}\email{haktas@ohu.edu.tr}

\affil*[1]{\orgdiv{Electrical and Computer Engineering Department}, \orgname{Drexel University}, \orgaddress{\city{Philadelphia}, \country{USA}}}

\affil[2]{\orgdiv{Electrical and Computer Engineering Department}, \orgname{A\&M University}, \orgaddress{\city{Prairie View}, \country{USA}}}

\affil[3]{\orgdiv{Electrical and Computer Engineering Department}, \orgname{Omer Halisdemir University}, \orgaddress{\city{Nigde}, \country{TR}}}


\abstract{Reconfigurable architectures like Field Programmable Gate Arrays (FPGAs) have been used for accelerating computations in several domains because of their unique combination of flexibility, performance, and power efficiency. However, FPGAs have not been widely used for high-performance computing, primarily because of their programming complexity and difficulties in optimizing performance. We optimize Tensil AI's open-source inference accelerator for maximum performance using ResNet20 trained on CIFAR in this paper in order to gain insight into the use of FPGAs for high-performance computing. In this paper, we show how improving hardware design, using Xilinx Ultra RAM, and using advanced compiler strategies can lead to improved inference performance. We also demonstrate that running the CIFAR test data set shows very little accuracy drop when rounding down from the original 32-bit floating point. The heterogeneous computing model in our platform allows us to achieve a frame rate of 293.58 frames per second (FPS) and a \%90 accuracy on a ResNet20 trained using CIFAR.  The experimental results show that the proposed accelerator achieves a throughput of 21.12 Giga-Operations Per Second (GOP/s) with a 5.21 W on-chip power consumption at 100 MHz. The comparison results with off-the-shelf devices and recent state-of-the-art implementations illustrate that the proposed accelerator has obvious advantages in terms of energy efficiency.}




\maketitle

\section{Introduction}\label{sec1}

Real-time vision-based motion tracking is necessary for many applications. Real-time video streaming for surveillance applications requires advanced encoding and decoding techniques as well as compute-intensive image processing designs. The ability to operate in real-time is especially important for applications in which speed is paramount, such as production areas, traffic speed control systems, or when the camera activity needs to be synchronized with other system components. The developers of machine vision frameworks and respectability may be engrossed in determining which of these steps to implement while developing the rest of the framework. The organize option is commonly selected when prototyping the system for the first time. The number of sections or outlines the application must prepare each instant depends on how many sections the prototyped application must handle at any given moment. Researchers are developing methods to design embedded systems that require less power, which is critical for most applications in modern embedded systems \cite{wang2019convolutional} \cite{vanderbauwhede2013high}.
High-resolution image preparation applications demand faster, configurable, high-throughput frameworks with superior productivity for preparing enormous data sets \cite{blaiech2019survey} \cite{zou2013high} \cite{isik2023energy}. FPGAs (Field-Programmable Gate Arrays) can play an important role since they provide configurability, adaptability, and parallelism to coordinate the necessary throughput rates of the application under consideration~\cite{woods2008fpga}. An FPGA device provides an execution method that allows it to be used in real-life applications. FPGAs have significantly increased the flexibility of hardware in general. A wider community of builders can now make use of these devices thanks to advancements in the toolchains for developing applications on them. Applications that require concurrency, high transfer speeds, and re-programmability typically use FPGAs. Modern digital life is increasingly reliant on image-processing applications. They are used in a variety of applications, including medical imaging, security, autonomous vehicles, and entertainment. In order to meet the increasing demand for more accurate and faster image processing, high-performance computing systems are needed. Image processing systems can be improved through FPGA-based design optimization. There are several factors that require pushing for higher performance in image processing. The following factors are discussed in more detail.
\begin{itemize}
  
\item Resolution and Image Size:
An image processing system's performance is strongly influenced by image resolution and file size. The complexity of the image processing algorithms required to analyze images increases with increasing resolution and size. For example, medical images such as CT scans and MRI images can have resolutions of several megapixels, and the files can be several gigabytes in size. Images of this size and complexity require high-performance computing systems that can handle large amounts of data quickly and accurately.

\item Real-Time Processing:
Real-time image processing is often required by image processing applications. Video streaming, security systems, and autonomous vehicles are applications in which real-time processing is essential. Detecting and avoiding obstacles, pedestrians, and other vehicles requires real-time image processing in autonomous vehicles. In order for these applications to operate smoothly, high-performance computing systems must be able to process large volumes of data in real-time.

\item Complex Algorithms:
High performance is also necessary due to the complexity of image processing algorithms. Complex algorithms require more processing power and memory to run. In image processing applications such as object recognition and image classification, deep learning algorithms require a significant amount of processing power and memory. High-performance computing systems accelerate the execution of these complex algorithms, providing faster and more accurate results.

\item Parallel Processing:
Parallel processing can greatly benefit image processing applications since multiple computations can be performed at once. Parallel processing is made possible through FPGA-based design optimization since FPGAs offer a high degree of parallelism. Multi-pixel image processing is possible with FPGAs, which allows for faster and more efficient image processing. The importance of this is especially pronounced in applications such as video processing where multiple frames must be processed simultaneously.
\end{itemize}
The increasing demand for faster and more accurate image processing requires image processing applications to push for higher performance. High-performance computing systems are needed because of factors such as high resolution and image size, real-time processing, complex algorithms, and parallel processing. Optimising FPGA-based designs for image processing applications is a very effective way to increase performance, and it is likely to become more relevant as the demand for faster and more accurate image processing increases. FPGAs are integrated circuits that can be programmed and reprogrammed to perform specific tasks. The unique features that make them well-suited to high-performance computing make them increasingly popular. High-performance computing can benefit from FPGAs as outlined below.
\begin{itemize}

\item High Parallelism:
Parallelism is one of the key features of FPGAs that makes them well-suited to high-performance computing. An FPGA can execute multiple tasks or operations simultaneously, which is essential for high-performance computing applications requiring a large amount of processing power. FPGAs achieve parallelism by using configurable logic blocks that can perform different tasks simultaneously.

\item Customizable Architecture:
The flexible architecture of FPGAs makes them ideal for high-performance computing applications. FPGAs can be programmed to meet specific performance requirements, enabling them to optimize performance for particular applications. Consequently, FPGAs can be customized to meet the specific needs of high-performance computing applications, something that isn't possible with general-purpose processors.

\item Low Latency:
High-performance computing can also be achieved with FPGAs due to their low latency. System latency is the amount of time it takes for an input to be processed. An FPGA can process inputs in nanoseconds, which is much faster than a general-purpose processor. Real-time applications, such as video and audio processing, require low latency to avoid poor quality.

\item High Bandwidth:
FPGAs have high bandwidth, which refers to the amount of data they can transfer. The ability to transfer large amounts of data quickly is an important feature for high-performance computing applications. Through high-speed serial transceivers, FPGAs can achieve high bandwidth that can reach several gigabits per second.

\item Energy Efficiency:
High-performance computing is also made possible by FPGAs' energy efficiency. Compared with general-purpose processors, FPGAs have lower power consumption, which is important for applications requiring a high level of processing power. Due to their parallel architecture, FPGAs achieve high energy efficiency and can be customized to meet application requirements.
\end{itemize}
FPGAs are an attractive option for applications requiring a significant amount of processing power, such as image processing, machine learning, and real-time processing. FPGAs are likely to become even more important in high-performance computing as the demand grows.

Tensil AI creates hardware and software solutions for machine learning applications. They offer high-performance machine learning inference on FPGA platforms through an open-source inference accelerator. As an open-source machine learning library developed by Google, TensorFlow Lite Inference Engine underpins Tensil AI's inference accelerator. The Tensil AI accelerator can therefore be easily integrated with existing machine learning applications. The Tensil AI inference accelerator performs quantization as one of its key features. A quantification process reduces the precision of machine learning models, making them easier to store and deploy. A Tensil AI accelerator performs quantization on the fly, which reduces the memory and power requirements of the inference engine. Its ability to support dynamic shapes is another key feature of the Tensil AI inference accelerator. Machine learning applications that require real-time processing of sensor or camera data can benefit from variable input data sizes and shapes. The Tensil AI accelerator is able to change the size of the inference engine on the fly based on the size of the input data, so it can handle a wide range of input sizes and shapes. Tensil AI inference accelerators are highly configurable, allowing developers to optimize their performance according to their needs. The low latency, high bandwidth, and high throughput processing it provides make it an ideal solution for high-performance computing applications. The Tensil AI inference accelerator is not only highly scalable but also highly efficient. The technology can be employed in edge devices such as smartphones, smart cameras, and IoT devices, as well as in cloud-based applications that require high-performance machine learning inferences. Tensil AI accelerators can be deployed on a wide range of FPGA platforms, including Xilinx's Alveo accelerator cards, making them ideal for high-performance computing applications. The Tensil AI open-source inference accelerator is a powerful tool for accelerating machine learning inference on FPGA platforms. A wide range of input sizes and shapes can be supported, making it a highly scalable and versatile solution. High-performance computing will likely become even more reliant on solutions like the Tensil AI inference accelerator as machine learning becomes more important \cite{WinNT} \cite{WinNT1}.

\vspace{.5cm}The rest of the paper is organized as follows: \textbf{Section II} presents the motivation and related works. \textbf{Section III} introduces open-source ml inference accelerators. The proposed method and its experimental results and analysis are reported in \textbf{Section IV} and \textbf{Section V}. \textbf{Section VI} concludes the contents of this paper and gives future aspects of this paper.

\section{Motivation}
FPGAs have been around for several decades, and they are used in many different applications. High-performance computing has been limited by a number of challenges and difficulties. FPGAs have not been widely used in high-performance computing due to their high development cost and complexity. The tools and technologies required for FPGA development are often expensive and complex, which makes it difficult to develop systems based on FPGAs. FPGA-based solutions have proven challenging to adopt for many organizations, especially for smaller organizations or those with limited resources. The limited availability of high-level software tools is another challenge with FPGAs in high-performance computing. Developing software for FPGAs requires a deep understanding of the underlying hardware architecture, which is more difficult than for traditional processors. However, high-level synthesis tools are not as mature as those used for traditional processors, making development more challenging \cite{sundararajan2010high} \cite{sklyarov2019hardware}.

Some high-performance computing applications can also be limited by the limited amount of on-chip memory on FPGAs. There is a significant amount of data transfer between the FPGA and external memory, which slows performance and increases latency. For many high-performance computing applications, floating-point operations are also not supported by FPGAs.  FPGAs used in high-performance computing also have a limited number of prebuilt IP blocks. The development of FPGA-based solutions often requires the use of pre-built intellectual property (IP) blocks, such as memory controllers and data interfaces. The availability of these IP blocks for FPGAs is often limited, which makes developing FPGA-based systems more difficult and time-consuming.

High-performance computing applications benefit from the advantages of FPGAs, despite these challenges. FPGAs can be highly optimized for specific tasks and often perform better than traditional processors in specific applications. A hardware-level parallelism capability also enhances performance for certain tasks. Recent developments have made FPGAs more accessible for high-performance computing, thus addressing these challenges. The availability of high-level synthesis tools for FPGAs makes software development easier, for example. A number of pre-built IP blocks are also being developed and made available for FPGAs. A number of FPGA-based solutions are now available that require less specialized hardware design knowledge and are easier to use. Despite the challenges and difficulties involved in developing and implementing FPGAs, they have not been widely used for high-performance computing, but efforts are being made to resolve these issues and make FPGA-based solutions more accessible and usable for high-performance computing. The adoption of FPGAs in high-performance computing will increase as development tools, IP blocks, and FPGA-based solutions improve \cite{huang2019accelerating}.

High-performance computing applications have attracted significant interest in FPGAs in recent years. FPGA-based systems can be highly optimized for specific tasks, and they can often perform better than traditional processors in specific applications.  The image and video processing industry has extensively used FPGAs for high-performance computing. The processing of high-resolution images and video can be carried out in real time using FPGAs.  A high-level synthesis tool called Vivado HLS has been used by researchers at UCLA to develop an FPGA-based system for real-time image processing \cite{chen2022fpga}. A throughput of 52 frames per second was achieved when filtering images, and 20 frames per second when segmenting images. High-performance computing has also been done using FPGAs in the financial industry. Complex mathematical operations are often involved in financial calculations, which are well suited for FPGAs. A high-frequency trading system developed by the Tokyo Stock Exchange (TSE) can process trades in less than one microsecond using FPGAs \cite{kohda2021characteristics} \cite{kohda2022characteristics}. The system uses FPGAs to calculate financial instruments such as options and futures. Machine learning and artificial intelligence are other areas where FPGAs have been used for high-performance computing. FPGAs can be highly optimized for neural network computations, making it possible to process large amounts of data faster and more efficiently. Scientific calculations can be highly optimized on FPGAs, resulting in faster, more efficient processing of large amounts of data. Furthermore, a number of existing works focus on optimizing FPGA-based systems for high-performance computing in general. Researchers have developed a tool called FireSim to simulate large-scale FPGA-based systems using cloud resources \cite{karandikar2019using}. The tool can be used to optimize system performance and evaluate different design options. There are many existing works that focus on using FPGAs for high-performance computing. Several applications, including image and video processing, finance, machine learning, artificial intelligence, and scientific research, have been demonstrated using FPGAs in these studies. With the continued development of FPGA-based tools and technologies, we can expect to see even increased adoption of FPGAs for high-performance computing in the future.

\section{Open-source ML inference accelerators}
Many high-performance computing applications rely on machine learning inference. Models are used to analyze input data and generate output results. High-performance computing systems can help speed up ML inference, which is often computationally intensive. High-performance computing applications may benefit from open-source ML inference accelerators. An ML inference accelerator is a specialized hardware device that performs ML inference tasks efficiently. It is typically performed on general-purpose processors or graphics processing units (GPUs) that are not optimized for ML inference. An ML inference accelerator provides efficient and optimized execution of ML inference tasks. Open-source ML inference accelerators offer the advantage of being free and customizable to fit specific use cases. An open-source ML inference accelerator offers a transparent and open development process, which encourages community participation and feedback. Open-source accelerators can also reduce the cost and time associated with developing ML inference accelerators. Recently, the Versatile Tensor Accelerator (VTA) has gained significant attention as an open-source ML inference accelerator. Inference tasks using VTA are performed with the help of a highly optimized hardware accelerator. TensorFlow, PyTorch, and ONNX are among the popular ML frameworks supported by VTA. There are a variety of hardware platforms that can be used with VTA, including FPGAs and ASICs \cite{moreau2019hardware}.

\begin{figure}[h!]
    \centering
    \includegraphics[width=0.8\textwidth]{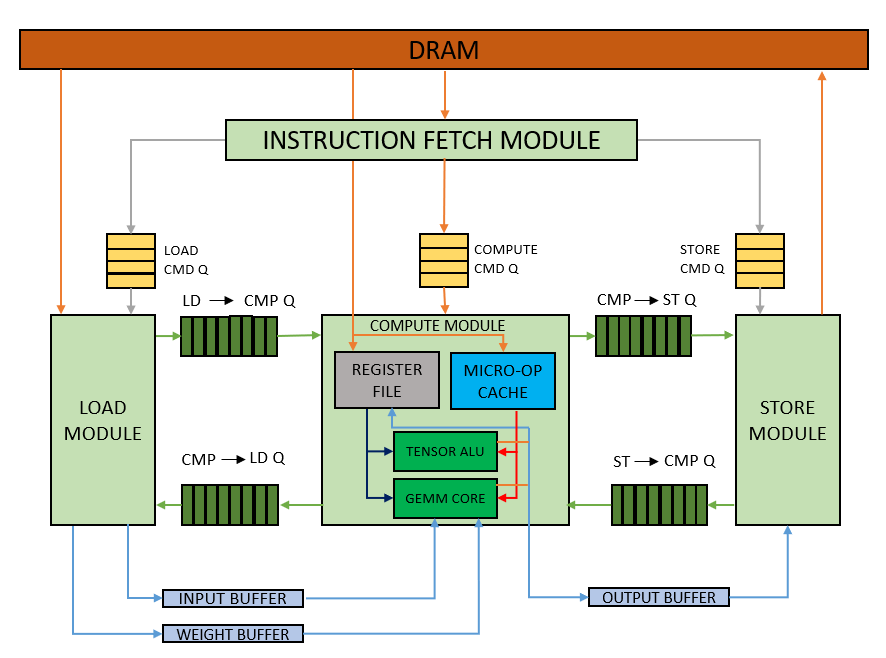}
    \caption{VTA Framework}\cite{moreau2019hardware}
    \label{fig_7}
\end{figure}

By providing open-source tools and platforms, developers can collaborate to create new and more efficient ML inference solutions. Collaboration can lead to faster development and adoption of new ML techniques and applications. Other open-source ML inference accelerators, such as Intel's OpenVINO and Xilinx's Deep Learning Processor, are available alongside VTA \cite{zunin2021intel} \cite{WinNT2}. These accelerators provide developers with a variety of options for building and optimizing machine learning systems. High-performance computing applications can take advantage of open-source ML inference accelerators because they provide a flexible and powerful tool. Development of ML inference systems can be customized and optimized, resulting in lower development costs and time, collaboration and innovation, and wide adoption of new ML techniques and applications. As the field of ML continues to grow and evolve, we can expect to see even more powerful and efficient open-source ML inference accelerators become available. 

High-performance computing applications using FPGAs can be built with Nengo and Tensil AI frameworks. A variety of hardware platforms, including FPGAs, can be used to build large-scale neural models with Nengo software. In addition, Nengo is designed to be flexible and extensible, allowing users to create customized models and algorithms. Nengo is well-suited for applications such as robotics and cognitive modeling because it can build complex models with many neurons and synapses. The Tensil AI hardware accelerator, on the other hand, performs machine learning inference tasks. The Tensil AI is designed to provide high performance at low power consumption, making it ideal for applications such as image recognition and natural language processing. Tensil AI is designed to be easily integrated with existing hardware architectures and supports a wide range of machine learning frameworks, including TensorFlow and PyTorch. Their focus is one of the key differences between Nengo and Tensil AI. Tensil AI focuses on accelerating machine learning inference tasks, whereas Nengo is primarily focused on building large-scale neural networks \cite{morcos2019nengofpga} \cite{dewolf2020nengo} \cite{gosmann2017automatic}.  

Nengo is more versatile and can be used for a wider variety of tasks, whereas Tensil AI is more focused on particular tasks. Their development processes are also key differences between Nengo and Tensil AI. The Open-source project Nengo is actively developed and maintained by a large developer community. As a result, users have access to a variety of resources, including documentation, tutorials, and support forums. The Tensil AI product, on the other hand, is a commercial product developed and supported by Tensil AI. Due to this, users have access to dedicated support and resources, but not as much community support as with an open-source project. Machine learning inference tasks can be performed quickly and with low latency using Tensil AI. Self-driving cars and industrial automation, for example, can benefit from their ability to make inferences quickly and efficiently.  Nengo, on the other hand, simulates complex behaviors over long periods of time using large-scale neural models. Tensil AI has the potential drawback of limited flexibility. Its limited versatility may be due to its specialized nature as a hardware accelerator. Users may not be able to create custom models or algorithms, and they may have to use pre-built models and architectures instead. Therefore, Nengo and Tensil AI are both powerful frameworks for developing high-performance computing applications. A variety of applications can be carried out with Nengo, whereas Tensil AI is more suited for specific tasks, such as machine learning inference. Developers should carefully evaluate the strengths and weaknesses of each framework before selecting one, and ultimately their choice will depend on the specific needs of the application \cite{WinNT} \cite{WinNT1}.

\section{Method}
We propose different approaches such as Vivado hardware design, leveraging Xilinx Ultra RAM, and using advanced compiler strategies to improve the performance of inference. In the ResNet20-ZCU104 tutorial by Tensil AI, several methods are used to optimize the design of the ResNet20 neural network for implementation on the Xilinx ZCU104 development board using their open-source inference accelerator. ResNet-20 is trained using CIFAR-10, a dataset that contains 60,000 32x32 RGB images categorized into ten classes. A PyTorch framework is used to train the network, which has an accuracy of approximately \%91. Training the network on ZCU104 is followed by several optimization steps for deployment on the device. A pruning technique reduces computation and memory requirements by removing unnecessary connections from the network. Tensil AI reduces the number of parameters and computations by pruning connections from the trained network. A further optimization method used in the tutorial is quantization, which involves reducing the weights and activations of the network. The network is quantized to 8-bit fixed-point precision using the TensorRT framework, further reducing memory and computation requirements. Tensil AI's open-source inference accelerator, designed to accelerate sparse neural network execution on FPGAs, implements the optimized neural network on the ZCU104. High performance and energy efficiency are achieved by utilizing FPGAs' reconfigurability and parallelism. The ResNet20-ZCU104 tutorial by Tensil AI demonstrates a variety of optimization techniques that can be used to optimize neural network designs for implementation on FPGA-based accelerators, including pruning and quantization.
\subsection{Baseline design}

Specifying 32 by 32 systolic array size contributed to the high utilization of multiply-accumulate units (DSP). Note how we pushed Block RAM (BRAM) utilization almost to its limit by specifying 16 KV local memory and 4 KV accumulators (KV = 1024 vectors = 1024 * 32 * 16 bits). The ZCU104 board supports an SD card interface. This allows us to use Tensil embedded driver file system functionality to read the ResNet model and a set of images to test it with. The set we will be using is the test set for the original CIFAR-10. The ResNet model is trained with separate training and validation sets from the CIFAR-10. The test set is what the model hasn’t seen in training and therefore gives an objective estimate of its accuracy. The CIFAR-10 provides a test set of 10,000 images in several formats. We will use the binary format that is more suitable for the embedded application. With the SD card inserted and containing the CIFAR-10 test data set and the ResNet model compiled for Tensil, you should see the inference printing every 100’s images and the corresponding prediction along with measured inferences (frames) per second. After running the inference on the entire test data set the program will print the final average frames per second and the accuracy of the inference. For the baseline solution, we are getting an average of 133.54 frames per second with \%90 accuracies. Note that the accuracy we are seeing when testing the same ResNet model with TensorFlow is \%92. The \%2 drop is due to changing the data type from a 32-bit floating point in TensorFlow to a 16-bit fixed point in Tensil.

\subsection{Dual clock solution}
\begin{figure}[h!]
    \centering
    \includegraphics[width=0.7\textwidth]{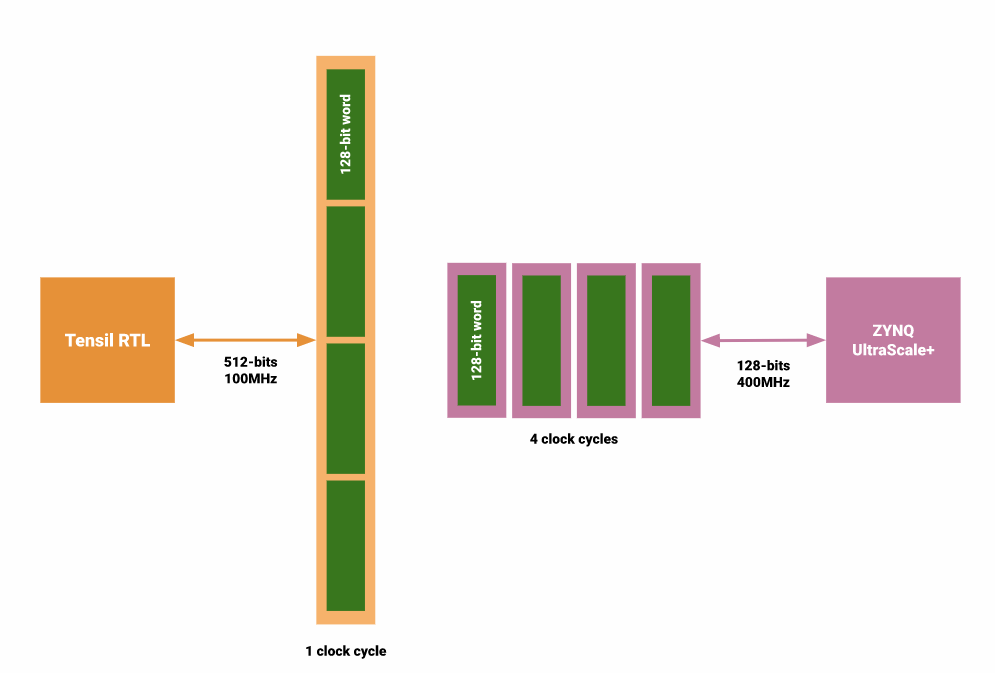}
    \caption{Tensil RTL clock domain. }
    \label{fig_2}
\end{figure}

The first optimization is based on the following observation. The Tensil RTL block is clocked at 100MHz. The Tensil block DRAM0 and DRAM1 ports are connected to AXI interfaces on the ZYNQ block. The instruction port is indirectly connected to the AXI on the ZYNQ block via AXI DMA. ZYNQ UltraScal+ AXI ports support up to 333MHz and a maximum width of 128 bits. This gives us the opportunity to introduce a second clock domain for 333MHz while at the same time making the Tensil AXI ports wider. Figure \ref{fig_2} shows how this may work in a simpler 100MHz to 400MHz, 512- to 128-bit conversion. Each clock in the Tensil clock domain would pump one 512-bit word in or out. This would match 4 clocks in the ZYNQ clock domain with 512-bit word split to or composed from 4 128-bit words.

For the dual clock solution, we are getting an average of 152.04 frames per second–a meaningful improvement over the baseline. This improvement is roughly proportional to the ratio of time spent in moving data to and from the FPGA to the time spent in internal data movement and computation. An accelerator is designed in two parts - a data path and a control path. Data paths process input data through neural networks, while control paths manage data flow and control the overall operation of the accelerator. Data and control paths are clocked separately at different frequencies, with the data path clocked at a higher frequency to maximize the accelerator's throughput. Data synchronization and timing violations between the two paths can also impact the accelerator's performance and reliability with this approach. Tensil AI's dual clock solution includes a number of design techniques such as pipelining, synchronization signals, and careful timing analysis to ensure proper data synchronization and avoid timing violations. A high level of throughput is still maintained while these techniques improve the accelerator's performance and reliability.

\subsection{Ultra RAM solution}
Ultra RAM refers to a design approach that optimizes memory access and utilization in FPGA-based inference accelerators. An Ultra RAM is a high-density memory block that is available in Xilinx FPGAs and offers high bandwidth and low latency access to memory. The Ultra RAM solution is used in the ResNet20-ZCU104 project to store the weights of the neural network model, which is a critical part of inference. Ultra RAMs are used to store weights so they can be accessed quickly and efficiently during the inference process. As part of the Ultra RAM configuration, the accelerator also supports concurrent reads and writes, further improving performance. Tensil AI's design approach makes optimal use of Ultra RAMs by using techniques such as weight compression and quantization, which reduce the memory footprint of weights without compromising accuracy. Using these techniques increases the capacity and efficiency of Ultra RAMs, improving the accelerator's performance overall. Inference accelerators based on FPGAs benefit greatly from the Ultra RAM solution for optimizing memory access and utilization. A ResNet20-ZCU104 neural network model has been successfully inferred with high performance and efficiency using the ResNet20-ZCU104.

The second optimization is based on the higher-end ZYNQ UltraScale+ device's support for another type of on-chip memory called Ultra RAM. By default, Vivado maps dual-port memory to Block RAM. In order for it to map to the Ultra RAM it needs hints in the Verilog code. To enable these hints we will use the Xilinx ultra ram option of the Tensil RTL tool. The amount of Ultra RAM available on ZCU104 allows us to add around 48 KV memory in addition to 20 KV available through Block RAM. We start by creating a new Tensil architecture for ZCU104 in which we allocate all of the Block RAM (20 KV) to accumulators and all of the Ultra RAM (48 KV) to local memory. For the Ultra RAM solution, we are getting an average of 170.16 frames per second, another meaningful improvement. This improvement is based purely on having larger on-chip memory. With a small on-chip memory the Tensil compiler is forced to partition ResNet convolution layers into multiple load-compute-save blocks. This, in turn, requires that the same input activations are loaded multiple times, assuming weights are loaded only once. This is called weight-stationary dataflow. In the future, we will add an option for input-stationary dataflow. With it, when partitioned, the input activations are loaded once and the same weights are loaded multiple times.FPGA utilization for Ultra RAM design is shown in Table \ref{table:table_1}.

\begin{table}
\centering
    \begin{tabular}{| c | c |} 
     \hline
     Utilization &XCZU7EV \\ 
     \hline\hline
    
     \hline
    LUT & 181440  \\ 
   
     \hline
   DSP & 1054 \\ 
     \hline
     BRAM & 293  \\ 
     \hline
    URAM & 96  \\ 
     
     \hline
    \end{tabular}
\caption{Resource Usage.}
\label{table:table_1}
\end{table}

\begin{figure}[h!]
    \centering
    \includegraphics[width=0.7\textwidth]{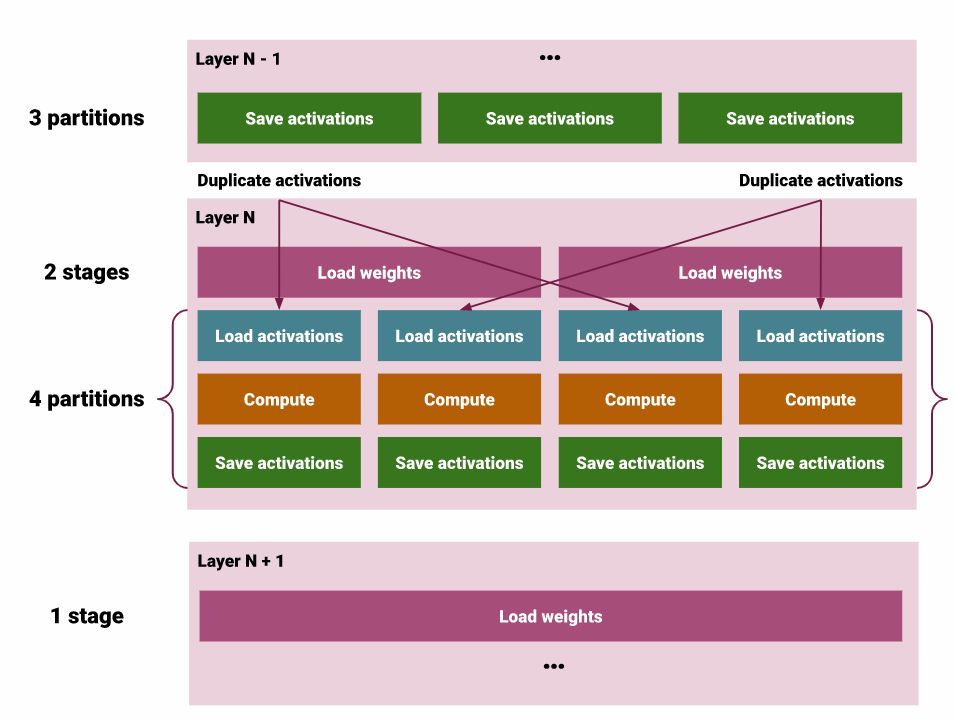}
    \caption{3-Partitioned Compilation.}
    \label{fig_3}
\end{figure}

Figure \ref{fig_3} shows such a 3-partitioned compilation. Layer N has 2 stages. In each stage, a unique subset of weights is loaded. Then, each stage is further split into 2 partitions. Partition is defined by the largest amount of weights, input and output activations, and intermediate results that fit local memory and accumulators.

\begin{figure}[h!]
    \centering
    \includegraphics[width=0.7\textwidth]{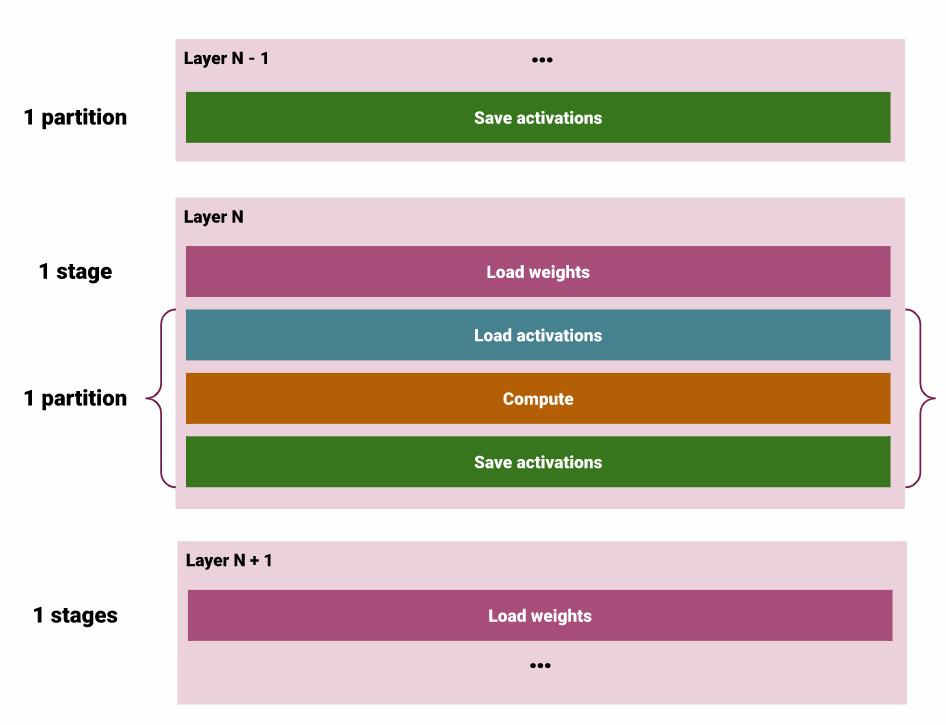}
    \caption{1-Partitioned Compilation.}
    \label{fig_4}
\end{figure}

Having larger on-chip memory reduces this partitioning and, by extension, the need to load the same data multiple times. Figure \ref{fig_4} shows how to layer N now has 1 stage and 1 partition that fits larger local memory and accumulators, which allows weights and activations to be loaded only once.

\subsection{Compiler Strategy with large local memory}

\begin{figure}[h!]
    \centering
    \includegraphics[width=0.7\textwidth]{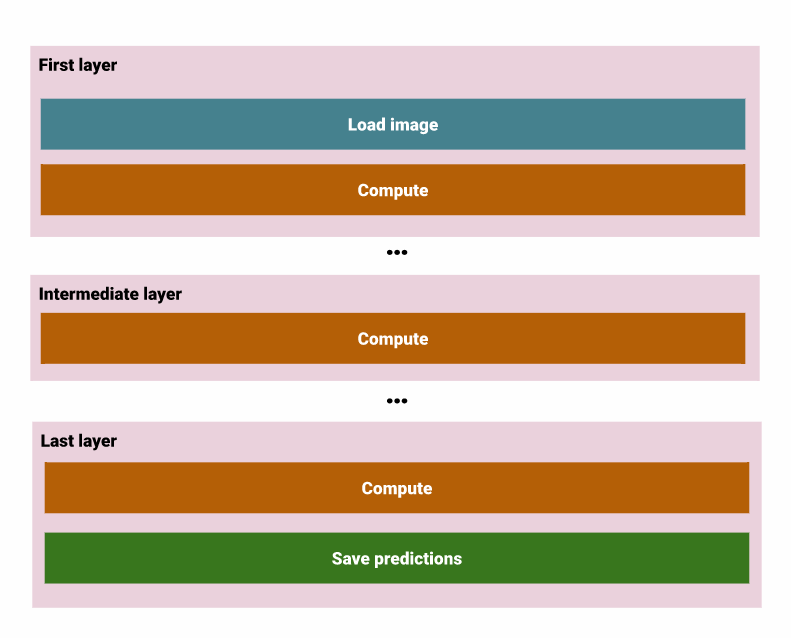}
    \caption{Compiler Strategy.}
    \label{fig_5}
\end{figure}

This strategy involves optimizing the speed and latency of local memory resources in FPGAs, such as block RAMs and Ultra RAMs, which are faster than external memory resources such as DRAM. Inference is carried out using the data and weights stored in the local memory, which are used for data storage and weight storage. Several techniques are used by ResNet20-ZCU104 to implement the compiler strategy with large local memory, such as weight compression and quantization, which reduce the memory footprint of weights without compromising their accuracy. The reduced memory footprint allows for larger portions of the neural network model and data to be stored in the local memory resources. The final optimization is based on the same hardware design and Tensil architecture we created to support the Ultra RAM. We will only change the Tensil compiler strategy. Tensil compilers, by default, assume that the model is much larger than the FPGA's local memory in terms of its weights and activations. This is true for large models and for low-end FPGA devices. For small and medium-sized models running on large FPGA devices, there is a possibility that local memory is large enough to contain the weights plus input and output activations for each layer. Our Proposed compiler strategy is shown in Figure \ref{fig_5}.

\begin{sidewaystable}
\sidewaystablefn%
\centering
\caption{Evaluation results on CPU, GPU, and our processor.}
\label{table:table_2}
\begin{tabular}{|c|c|c|c|c|c|c|}
\hline
&\begin{tabular}[c]{@{}c@{}}Technology\\ {[}nm{]}\end{tabular} & \begin{tabular}[c]{@{}c@{}}Frequency\\ {[}MHz{]}\end{tabular}& \begin{tabular}[c]{@{}c@{}}Latency\\ {[}ms{]}\end{tabular} & \begin{tabular}[c]{@{}c@{}}Throughput\\ {[}GOP/s{]}\end{tabular}  &\begin{tabular}[c]{@{}c@{}} Power {[}Watt{]}\end{tabular}& \begin{tabular}[c]{@{}c@{}}Energy Efficiency\\ {[}GOP/s/W{]}\end{tabular} \\ \hline
Intel Xeon E5-2697 (CPU) & 14 & 2300 & 1137.62 & 27.20&145 &0.19 \\ \hline
NVIDIA GTX 1080 TI (GPU) & 14 & 1481 & 6.15 & 235.77&250 &0.94\\ \hline
Xilinx ZCU104 (FPGA) & 16 & 100 & 2.91 & 21.12&5.21 &4.05\\ \hline

\end{tabular}
\end{sidewaystable}

\section{Results}
Our results demonstrate the effectiveness of Tensil AI's open-source inference accelerator for optimizing neural networks and implementing them on FPGAs for high-performance computing applications. It has been done using CPUs, GPUs, and FPGAs. CPU/GPU-based NNs consume a lot of power and have a limited throughput due to limited memory bandwidth which is shown in Table \ref{table:table_2}. In Table \ref{table:table_3} Many researchers have developed FPGA-based designs for accelerating network inference workloads in order to achieve better energy efficiency.FPGAs function as programmable devices that can construct unique logic, alleviating constraints on neural network implementation. We demonstrated how improving the Vivado hardware design, leveraging Xilinx Ultra RAM, and using advanced compiler strategies can improve the performance of inference. As a result, one of the current research hotspots involves the development of hardware systems supporting NN inference based on FPGA to achieve high throughput and power efficiency. 

\begin{figure}[h!]
    \centering
    \includegraphics[width=0.7\textwidth]{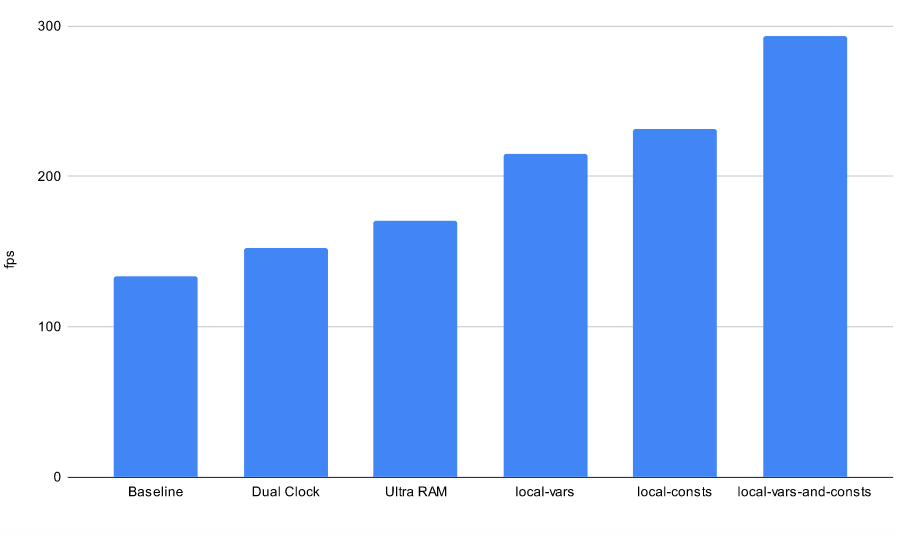}
    \caption{Performance Chart.}
    \label{fig_6}
\end{figure}

Figure \ref{fig_6} summarizes presented solutions and their frames per second performance.

\begin{table*}
    \renewcommand{\arraystretch}{1.2}
    	\setlength{\tabcolsep}{3pt}
    \centering
    \caption{Comparisons with previous implementations.}
    \label{table:table_3}
    \resizebox{0.8\linewidth}{!}{
    \begin{tabular}{llllllll}
         Work&Device  &Frequency &Quant  &Power& FPS & Throughput &Energy Efficiency  \\\hline
         {Ma et al.~\cite{ma2017optimizing}}&Arria-10 GX
    &$150MHz$&8-16 bit fixed&$21.2W$&$-
        $&$645.25(GOP/s)$&$30.44(GOP/s/W)$ \\
        {Mei et al.~\cite{mei2017200mhz}}&Virtex-7
   &$200MHz$&16-bit float&$10.81W$&$6.58$&$202.42
        (GOP/s)$&$1.64(GOP/s/W)$\\
        {Zhang et al.~\cite{zhang2019optimized}}&{Zynq ZU7EV}&\textbf{$300MHz$}&8-bit fixed&$17.67W$&$-$&$290.40
        (GOP/s)$&$0.80(GOP/s/W)$\\
        {Blott et al.~\cite{blott2018finn}}&{Zynq ZU3EG}&\textbf{$220MHz$}&8-bit fixed&$10.2W$&$200$&$400
        (GOP/s)$&$39.21(GOP/s/W)$\\
        {Zhang et al.~\cite{zhang2020efficient}}&{Virtex-7}&\textbf{$200MHz$}&8-bit fixed&$6.32W$&$6.77$&$209.60
        (GOP/s)$&$33.16(GOP/s/W)$\\
        {Li et al.~\cite{li2019efficient}}&{Zynq 7010}&\textbf{$200MHz$}&16-bit fixed&$19.52$&$-$&$452.8
        (GOP/s)$&$23.20(GOP/s/W)$\\
         {Suda et al.~\cite{suda2016throughput}}&{Stratix-V}&\textbf{$120MHz$}&8-16 bit fixed&$25.8W$&$-$&$117.8
        (GOP/s)$&$4.56(GOP/s/W)$\\
        {Ours}&{Zynq ZU7EV}&\textbf{$100MHz$}&32-bit floating&$5.21W$ &$290.58$&$21.12(GOP/s)$&$4.05(GOP/s/W)$\\
    \end{tabular}}
\end{table*}

\section{Conclusions}

The ResNet20-ZCU104 project demonstrates the potential of using FPGA-based acceleration for machine learning tasks. By leveraging the unique capabilities of FPGAs, such as low latency and efficient memory usage, the project achieved impressive results in terms of both performance and accuracy. The model is implemented for hardware acceleration with various heterogeneous devices and resulting in an energy-efficient, reconfigurable system on the latter. In the further phase of our work, we will propose to use Dynamic Partial Reconfiguration which is state-of-art technology of reconfigurable hardware into an achieved high-performance framework. Within this feature, we will solve reshaping and offloading the initial and post-data processing for high-performance computing with Tensil AI. The Tensil AI already can take a different model by compiling something new, but the data going in and out could require manipulation which works between the input process and output process of the Tensil AI.

\section{Declarations}

\subsection{Ethical Approval}
Not applicable

\subsection{Competing interests}
The authors declare no competing interests.

\subsection{Authors' contributions}
M.I., K.I., and H.A. contributed equally to the design, implementation, and evaluation of the FPGA-based inference accelerator for machine learning tasks presented in this paper. M.I. and K.I. performed the hardware design and optimization and the software development and performance evaluation, while H.A. contributed to a review of this paper. All authors contributed to the writing and revision of the manuscript and approved the final version for submission.

\subsection{Funding}
This research did not receive any specific grant from funding agencies in the public, commercial, or not-for-profit sectors.

\subsection{Availability of data and materials}
The data used in this study are available upon request.


\bibliography{sn-bibliography}


\end{document}